\documentclass[conference]{IEEEtran}

\ifCLASSINFOpdf
 
\else
  
\fi

\usepackage{amsmath,amssymb}
\usepackage{color}
\usepackage[noadjust]{cite}
\usepackage{graphicx}
\usepackage{epsfig}
\usepackage{breqn}
\usepackage[caption=false,font=footnotesize]{subfig}
\usepackage{array}

\newtheorem{theorem}{Theorem}[section]

\newtheorem{prop}[theorem]{Proposition}

\allowdisplaybreaks

\newcommand{\fsquare}{\vrule height6pt width7pt depth1pt}   

\newcommand{\myproof}{{\hfill \\ \bf Proof. \ }}           
\newcommand{\myendpf}{\hfill \fsquare \\ [0.1in]}             

\newcommand\given[1][]{\:#1\vert\:}

\usepackage[margin=0.75in]{geometry}

\begin{document}

\title{\vspace{0.25in} On the Network Reliability Problem of the Heterogeneous Key Predistribution Scheme}

\author{\IEEEauthorblockN{Rashad Eletreby and Osman Ya\u{g}an}
\IEEEauthorblockA{Department
of Electrical and Computer Engineering and CyLab, \\
Carnegie Mellon University, Pittsburgh,
PA, 15213 USA\\
reletreby@cmu.edu, oyagan@ece.cmu.edu}}


\maketitle

\begin{abstract}
We consider the network reliability problem in wireless sensor networks secured by the heterogeneous random key predistribution scheme. 
This scheme generalizes Eschenauer-Gligor scheme by considering the cases when the network comprises sensor nodes with varying level of resources; e.g., regular nodes vs. cluster heads. The scheme induces the {\em inhomogeneous} random key graph, denoted $\mathbb{G}(n;\pmb{\mu},\pmb{K},P)$.
We analyze the {\em reliability} of $\mathbb{G}(n;\pmb{\mu},\pmb{K},P)$ against random link failures. Namely, we consider $\mathbb{G}(n;\pmb{\mu},\pmb{K}, P,\alpha)$ formed by deleting each edge  of 
$\mathbb{G}(n;\pmb{\mu},\pmb{K},P)$
independently with probability $1-\alpha$,
and study the probability that the resulting graph i) has no isolated node; and ii) is connected. We present scaling conditions on $\pmb{K}$, $P$, and $\alpha$ such that both events take place with probability zero or one, respectively, as the number of nodes gets large. We present numerical results to support these in the finite-node regime.


\end{abstract}


\IEEEpeerreviewmaketitle
\section{Introduction}
\IEEEPARstart{W}{ireless} sensor networks (WSNs) consist of low-cost, low-power, small sensor nodes that are typically deployed randomly in large numbers, with application areas as diverse as military, health, environmental monitoring, etc \cite{Akyildiz_2002}.
In most cases, WSNs are deployed in hostile environments, e.g., battlefields, making it crucial to use cryptographic protection to secure sensor communications. 
Therefore, significant efforts have been devoted to 
developing methods for securing WSNs, and 
{\em random key predistribution schemes} have been widely accepted as feasible solutions in the face of the unique challenges of WSNs. Namely,
limited computational capabilities, limited transmission power, lack of a priori knowledge of deployment configuration, and vulnerability to node capture attacks; e.g., 
see \cite{Gligor_2002,Perrig_2004,Xiaojiang_2008,camtepe_2005} for a detailed discussion on security challenges in WSNs and solutions based on key predistribution.
In this paper, we consider a 
{\em heterogeneous} key predistribution scheme introduced recently by
Ya\u{g}an \cite{Yagan/Inhomogeneous} 
as a
variation of the {\em classical} Eschenauer-Gligor (EG) scheme \cite{Gligor_2002}.  The heterogeneous key predistribution scheme accounts for the cases when the network comprises sensor nodes with varying level of resources, e.g., regular nodes vs. cluster heads, which is likely to be the case for many WSN applications \cite{Yarvis_2005}. According to this scheme, each sensor belongs to one of $r$ priority classes that controls the {\em number} of cryptographic keys assigned to them. More specifically, each of the $n$ sensors is independently assigned to class-$i$ with probability $\mu_i>0$, for each $i=1,\ldots,r$; obviously we have $\sum_{i=1}^r \mu_i=1$. Sensors from  class-$i$ are each given $K_i$ keys selected uniformly at random from a pool of size $P$. Then, pairs of sensors that have at least one key in common can communicate securely after deployment. 
With $\pmb{\mu}=\{\mu_1, \ldots, \mu_r \}$ and $\pmb{K}=\{K_1,\ldots,K_r\}$, we let
$\mathbb{G}(n,\pmb{\mu},\pmb{K},P)$ denote the random graph
 induced by the heterogeneous key predistribution scheme.
This model was referred to as the {\em inhomogeneous} random key graph  in \cite{Yagan/Inhomogeneous}, wherein, zero-one laws for absence of isolated nodes and connectivity are established.


The main goal of this paper is to investigate the {\em reliability} of secure WSNs under the heterogeneous key predistribution scheme. In particular, to account for the possibility that links between two sensor nodes may fail (e.g., due to random failures, adversarial attacks, etc.), we apply 
a Bernoulli link-failure model to the inhomogeneous random key graph
$\mathbb{G}(n;\pmb{\mu},\pmb{K},P)$. Namely, we assume that each link 
in $\mathbb{G}(n;\pmb{\mu},\pmb{K},P)$ is operational with probability $\alpha$
and fails  with probability $1-\alpha$, independently from others.
This models random attacks as well as random failures 
due to sensor malfunctioning or harsh environmental conditions.

Let $\mathbb{G}(n;\pmb{\mu},\pmb{K},P,\alpha)$ denote the resulting random graph
that contains all operational links in $\mathbb{G}(n;\pmb{\mu},\pmb{K},P)$.
The network reliability problem is concerned \cite{net_rel}, \cite[Section 7.5]{Bollobas} with deriving
the probability that $\mathbb{G}(n;\pmb{\mu},\pmb{K},P,\alpha)$ exhibits certain desired properties -- that captures the ability of the network to continue its services -- as a function of the link failure probability $1-\alpha$. Here, we focus 
on two standard and related properties that the network i) has no {\em isolated} node, and ii) is connected. For arbitrary graphs with fixed size $n$, deriving these probabilities are known
\cite{Valiant,ProvanBall}
to be $\#P$-complete, meaning that no polynomial algorithm exists for their solution, unless $P=NP$. 
Given that it is not feasible to derive them, we study the {\em asymptotic} behavior of these probabilities as $n$ gets large, when the model parameters are scaled with $n$; the finite-node case is also considered via simulations.


Our contributions are as follows. We present conditions on how to scale
$\pmb{K},P$ and the link failure probability $\alpha$ such that 
the network $\mathbb{G}(n;\pmb{\mu},\pmb{K}_n,P_n,\alpha_n)$  is 
connected with probability approaching to one and zero, respectively, as $n$
grows unboundedly large. We establish an analogous zero-one law for $\mathbb{G}(n;\pmb{\mu},\pmb{K}_n,P_n,\alpha_n)$ to have no  node that is isolated (i.e., that has zero edge). 
These sharp results are likely to be useful in {\em dimensioning} the heterogeneous scheme, namely {\em controlling} the key ring parameters $(\pmb{\mu},\pmb{K})$, and key pool size $P$ such that the network has a desired level reliability against link failures. A particularly surprising conclusion derived from our results is that network reliability is tightly dependent on the {\em smallest} key ring size used in the network; see Section 
\ref{sec:results} for details.

Our results complement and generalize several previous work in the literature.
In particular, we complement the work \cite{Zhao_2014} that studies the reliability of secure WSNs against a fixed number $k$ of link failures; in our case the number of failed links can be unboundedly large. Our results also contain as special cases the zero-one laws for connectivity in inhomogeneous random key graphs
\cite{Yagan/Inhomogeneous} and reliability results in homogeneous random key graphs  \cite{Yagan/EG_intersecting_ER}; see 
Section \ref{sec:comparison} for details.


All limiting statements, including asymptotic equivalences are considered with the number of sensor nodes $n$ going to infinity. 
The indicator function of an event $E$ is denoted by $\pmb{1}[E]$. 
We say that an event holds with high probability (whp) if it holds with probability $1$ as $n \rightarrow \infty$. 
In comparing
the asymptotic behavior of the sequences $\{a_n\},\{b_n\}$,
we use the standard Landau notation, e.g.,
$a_n = o(b_n)$,  $a_n=w(b_n)$, $a_n = O(b_n)$, $a_n = \Omega(b_n)$, and
$a_n = \Theta(b_n)$.
We also use $a_n \sim b_n$ to denote the asymptotic equivalence $\lim_{n \to \infty} {a_n}/{b_n}=1$.

\section{The Model}
The heterogeneous random key predistribution scheme introduced
in \cite{Yagan/Inhomogeneous} works as follows. Consider a network of $n$ sensors labeled as $v_1, v_2, \ldots,v_n$. Each sensor node is classified into one of the $r$ classes, e.g., priority levels, according to a probability distribution $\pmb{\mu}=\{\mu_1,\mu_2,\ldots,\mu_r\}$ with $\mu_i >0$ for $i=1,\ldots,r$ and $\sum_{i=1}^r \mu_i=1$. Then, a class-$i$ node is assigned $K_i$ cryptographic keys selected uniformly at random and \textit{without replacement} from a key pool of size $P$. It follows that the key ring $\Sigma_x$ of node $v_x$ is a random variable (rv) with 
\begin{equation} \nonumber
\mathbb{P}[\Sigma_x=S \mid t_x=i]= \binom P{K_i}^{-1}, \quad S \in \mathcal{P}_{K_i},
\nonumber
\end{equation}
where $t_x$ denotes the class of $v_x$ and $\mathcal{P}_{K_i}$ is the collection of
all subsets of $\{1,\ldots, P\}$ with size $K_i$.
The classical key predistribution scheme of Eschenauer and Gligor \cite{Gligor_2002}
constitutes a special case of this model with $r=1$, i.e., when all sensors belong to the same class and receive the same number of keys; see also \cite{yagan2012zero,YM_ISIT2009}.

Let $\pmb{K}=\{K_1,K_2,\ldots,K_r\}$ and assume without loss of generality that $K_1 \leq K_2 \leq \ldots \leq K_r$. Consider a random graph $\mathbb{G}$ induced on the vertex set $\mathcal{V}=\{v_1,\ldots,v_n\}$ such that a pair of nodes $v_x$ and $v_y$ are adjacent, denoted by $v_x \sim_G v_y$, if they have at least one cryptographic key in common, i.e.,
\begin{equation}
v_x \sim_G v_y \quad \text{if} \quad \Sigma_x \cap \Sigma_y \neq \emptyset.
\label{adjacency_condition}
\end{equation}

The adjacency condition (\ref{adjacency_condition}) defines the inhomogeneous random key graph denoted by $\mathbb{G}(n;\pmb{\mu},\pmb{K},P)$  \cite{Yagan/Inhomogeneous}. This model is also known in the literature  as
the {\em general random intersection graph}; e.g., see \cite{Zhao_2014,Rybarczyk,Godehardt_2003}. 
 The probability $p_{ij}$ that a class-$i$ node and a class-$j$ node are adjacent is given by
\begin{equation}
p_{ij} = \mathbb{P}[v_x \sim_G v_y \mid t_x=i, t_y=j] =
1-\frac{\binom {P-K_i}{K_j}}{\binom {P}{K_j}}
\label{eq:osy_edge_prob_type_ij}
\end{equation}
as long as $K_i + K_j \leq P$; otherwise if $K_i +K_j > P$, we  have $p_{ij}=1$.
Let $\lambda_i$ denote the \textit{mean} probability that a class-$i$ node is connected to another node in $\mathbb{G}(n;\pmb{\mu},\pmb{K},P)$. We have
\begin{align}
\lambda_i & =\mathbb{P}[v_x \sim_G v_y \mid t_x=i ]
=\sum_{j=1}^r p_{ij} \mu_j.
 \label{eq:osy_mean_edge_prob_in_RKG}
\end{align}

To account for the possibility that links between two sensor nodes may fail, e.g., due to random failures, adversarial attacks, etc., we apply 
a Bernoulli link-failure model to the inhomogeneous random key graph
$\mathbb{G}(n;\pmb{\mu},\pmb{K},P)$: 
With $\alpha \in (0,1)$ let $\{B_{ij}(\alpha), 1 \leq i < j \leq n\}$ denote independent Bernoulli rvs, each with success probability $\alpha$. Then 
the link between sensors $v_x$ and $v_y$ is deemed to be operational (i.e., {\em up}) if $B_{xy}(\alpha)=1$, and not operational (i.e., {\em down}) if $B_{xy}(\alpha)=0$. Put differently, every edge in $\mathbb{G}(n;\pmb{\mu},\pmb{K},P)$ is deleted independently with probability $1-\alpha$\footnote{An interesting direction for future work would be to consider a heterogeneous link-failure model, where the link between a type-$i$ and type-$j$ node fails with probability $1-\alpha_{ij}$.}.

Let $\mathbb{G}(n;\pmb{\mu},\pmb{K},P,\alpha)$ denote the resulting random graph
that contains all the operational links in $\mathbb{G}(n;\pmb{\mu},\pmb{K},P)$.
To simplify  notation, we let $\pmb{\theta}=(\pmb{K},P)$, and $\pmb{\Theta}=(\pmb{\theta},\alpha)$. In $\mathbb{G}(n;\pmb{\mu},\pmb{\Theta})$,
distinct nodes $v_x$ and $v_y$ are adjacent, denoted $v_x \sim v_y$, 
if and only if they are adjacent in $\mathbb{G}(n;\pmb{\mu},\pmb{K},P)$ \textit{and} the edge $v_x \sim_{G} v_y$ is operational (i.e., has not failed).
By independence, the probability of an edge  between a class-$i$ node and a class-$j$ node in $\mathbb{ G}(n;\pmb{\mu},\pmb{\Theta})$ is then given by
\begin{equation} \nonumber
\mathbb{P}[v_x \sim v_y \mid t_x=i,t_y=j]=\alpha p_{ij}.
\end{equation}
Similar to (\ref{eq:osy_mean_edge_prob_in_RKG}), we denote the mean edge probability for a class-$i$ node in $\mathbb{ G}(n;\pmb{\mu},\pmb{\Theta})$ as $\Lambda_i$. It is clear that
\begin{align} 
\Lambda_i = \sum_{j=1}^r \mu_j \alpha p_{ij} = \alpha \lambda_i, \quad i=1,\ldots, r.
\label{eq:osy_mean_edge_prob_in_system}
\end{align}

Throughout, we assume that the number of classes $r$ is fixed and does not scale with $n$, and so are the probabilities $\mu_1, \ldots,\mu_r$. All other parameters are scaled with $n$.

\section{Main Results and Discussion}
We refer to a mapping $\pmb{\Theta}=K_1,\ldots,K_r,P,\alpha:
 \mathbb{N}_0 \rightarrow \mathbb{N}_0^{r+1} \times (0,1)$ as a \textit{scaling} 
if
\begin{equation}
1 \leq K_{1,n} \leq K_{2,n} \leq \ldots \leq K_{r,n} \leq P_n/2
\label{scaling_condition_K}
\end{equation}
for all $n=2,3,\ldots$. 
We note that under (\ref{scaling_condition_K}), the edge probability $p_{ij}$ is given by
(\ref{eq:osy_edge_prob_type_ij}).

\subsection{Results}
\label{sec:results}
We first present a zero-one law for the absence of isolated nodes in 
$\mathbb{ G}(n;\pmb{\mu},\pmb{\Theta}_n)$.
\begin{theorem}
\label{theorem:isolated_nodes}
Consider a probability distribution $\pmb{\mu}=\{\mu_1,\mu_2,\ldots,\mu_r\}$ with $\mu_i >0$ for $i=1,\ldots,r$ and a scaling 
$\pmb{\Theta}: \mathbb{N}_0 \rightarrow \mathbb{N}_0^{r+1} \times (0,1)$ such that
\begin{equation}
\Lambda_1(n)=\alpha_n \lambda_1(n) \sim c \frac{\log n}{n}
\label{scaling_condition_KG}
\end{equation}
for some $c>0$. We have
\begin{equation} \nonumber
\lim_{n\to\infty} \mathbb{P} \left[ \begin{split} \mathbb{G}(n;\pmb{\mu},\pmb{\Theta}_n) \text{ has} \\ \text{ no isolated nodes} \end{split} \right]=
\begin{cases} 
      \hfill 0    \hfill & \text{ if $c<1$} \\
      \hfill 1 \hfill & \text{ if $c>1$} \\
\end{cases}
\end{equation}
\end{theorem}
The scaling condition (\ref{scaling_condition_KG}) will often be used in the form
\begin{equation} \label{scaling_condition_KG_v2}
\Lambda_1(n)=c_n \frac{\log n}{n}, \ n=2,3,\ldots
\end{equation}
with $\lim_{n\to\infty} c_n=c>0$.

Next, we present an analogous result for connectivity.
\begin{theorem}
\label{theorem:connectivitiy}
Consider a probability distribution $\pmb{\mu}=\{\mu_1,\mu_2,\ldots,\mu_r\}$ with $\mu_i >0$ for $i=1,\ldots,r$ and a scaling $\pmb{\Theta}: \mathbb{N}_0 \rightarrow \mathbb{N}_0^{r+1} \times (0,1)$ such that
(\ref{scaling_condition_KG})
holds for some $c>0$. Then, we have
\begin{equation} \nonumber
\lim_{n\to\infty} \mathbb{P}[\mathbb{ G}(n;\pmb{\mu},\pmb{\Theta}_n) \text{ is connected}]=
\begin{cases} 
      \hfill 0    \hfill & \text{ if $c<1$} \\
      \hfill 1 \hfill & \text{ if $c>1$} \\
\end{cases}
\end{equation}
under the extra conditions that
\begin{equation}
P_n \geq \sigma n, \quad n=1, 2, \ldots
\label{eq:conn_Pn2}
\end{equation}
for some $\sigma>0$
and
\begin{equation}
\alpha_n p_{11}(n) =\omega\left(\frac{1}{n }\right).
\label{eq:conn_K1}
\end{equation}
\end{theorem}

Theorem~\ref{theorem:isolated_nodes} (resp. Theorem~\ref{theorem:connectivitiy}) states that $\mathbb{ G}(n;\pmb{\mu},\pmb{\Theta}_n)$ has no isolated node (resp. is connected) whp if the mean degree of class-$1$ nodes (that receive the smallest number $K_{1,n}$ of keys) is scaled as $(1+\epsilon) \log n$ for some $\epsilon > 0$. On the other hand, if this minimal mean degree scales as $(1-\epsilon) \log n$ for some $\epsilon > 0$, then whp $\mathbb{G}(n;\pmb{\mu},\pmb{\Theta}_n)$ has an isolated node, and hence not connected. 
These results indicate that the minimum key ring size in the network has a surprisingly
significant impact on the reliability of 
$\mathbb{G}(n;\pmb{\mu},\pmb{\theta}_n)$. This is more clearly seen under the additional assumption that $\lambda_1(n)=o(1)$ which gives \cite[Lemma 4.2]{Yagan/Inhomogeneous}
\[
\lambda_1(n) \sim \frac{K_{1,n} K_{\textrm{avg},n}}{P_n}
\]
where $K_{\textrm{avg},n}= \sum_{j=1}^r \mu_j K_{j,n}$ 
denotes the mean key ring size. Using this in (\ref{scaling_condition_KG}), we see that
for fixed mean number $K_{\textrm{avg},n}$ of keys per sensor, network reliability is directly affected by the minimum key ring size $K_{1,n}$. For example, reducing $K_{1,n}$ by half means that the smallest $\alpha$ for which the network remains connected is increased by two-fold, which then reduces the largest link failure probability $1-\alpha$ that can be sustained by a similar order.

The resemblance of the results presented in Theorem~\ref{theorem:isolated_nodes} and Theorem~\ref{theorem:connectivitiy} indicates that the absence of isolated nodes property and connectivity property are asymptotically equivalent for $\mathbb{ G}(n;\pmb{\mu},\pmb{\Theta}_n)$, similarly with
some well-known random graph models; e.g.,
 inhomogeneous random key graphs \cite{Yagan/Inhomogeneous}, ER graphs \cite{Bollobas}, and (homogeneous) random key graphs \cite{yagan2012zero}.


We remark that conditions (\ref{eq:conn_Pn2}) and (\ref{eq:conn_K1}) are enforced mainly for technical reasons and they are only needed in the proof of the one-law of Theorem~\ref{theorem:connectivitiy}. These conditions are likely to hold in real-world WSN implementations. In particular,  (\ref{eq:conn_Pn2})  should hold in practice to ensure the resiliency of the WSN against node capture attacks \cite{DiPietroTissec}, while (\ref{eq:conn_K1}) is needed as otherwise the 
network would be trivially disconnected \cite[Section 3.2]{Yagan/Inhomogeneous}.

%

\subsection{Comparison with related work}
\label{sec:comparison}

Our main results extend the work by Ya\u{g}an  \cite{Yagan/Inhomogeneous} who established zero-one laws for the connectivity of inhomogeneous random key graph $\mathbb{G}(n,\pmb{\mu},\pmb{K},P)$ without employing a link-failure model. It is clear that, although a crucial first step in the study of heterogeneous key predistribution schemes, the assumption that all links are operational, i.e., {\em reliable}, is not likely to hold in most practical settings. 
In this regard, our work extends the results by Ya\u{g}an \cite{Yagan/Inhomogeneous} to more practical WSN scenarios where the unreliability of links are taken into account. In fact, by setting  
$\alpha_n=1$ for each $n=1,2,\ldots$ (i.e., by assuming that all links are {\em reliable}), our results reduce to those given in
\cite{Yagan/Inhomogeneous}. 

The reliability of secure WSNs was also studied in \cite{Yagan/EG_intersecting_ER}, but under the Eschenauer-Gligor scheme \cite{Gligor_2002} where all sensors receive the same number of keys.
However, when the network consists of sensors with varying level of resources (e.g., computational, memory, power) and/or with varying level of security and connectivity requirements, it may no longer be sensible to assign the same number of keys to all sensors.
Our work addresses this issue by generalizing  \cite{Yagan/EG_intersecting_ER} to the cases where nodes can be assigned different number of keys. 
When $r=1$, i.e., when all nodes belong to the same class and receive the same number of keys, our
result recovers the main result in  \cite{Yagan/EG_intersecting_ER}.


%

Another notable work that is related to ours is by Zhao et al. \cite{Zhao_2014}, who studied the $k$-connectivity and $k$-robustness in the inhomogeneous random key graph. A graph is said to be $k$-connected if it remains connected after removal (i.e., \textit{failure}) of any $k-1$ nodes. Thus, the results obtained in \cite{Zhao_2014} ensure the reliability of the network against the failure of any $k-1$ nodes, for some integer constant $k$. Since $k$-vertex-connectivity implies $k$-edge-connectivity, the network is ensured to be reliable against the failure of at least $k-1$ edges, for some integer constant $k$. Our work complements these results by considering the case when \textit{each and every} edge fails with probability $1-\alpha$, 
so that the total number of failed links is possibly infinite;
e.g., as many as $O(n^2)$ links may fail.

\subsection{Significance of the results}
Our results are helpful in ensuring network reliability in multitude of applications where inhomogeneous random key graphs are utilized. For instance, reliability against the failure of wireless links is important in WSN applications where sensors are deployed in hostile environments (e.g., battlefield surveillance), or, are unattended for long periods of time (e.g., environmental monitoring), or, are used in life-critical applications (e.g., patient monitoring). 

Considering the asymptotic regime, a key question in network reliability analysis is whether or not there exists a threshold $\alpha_n^* \in (0,1)$ such that if $\alpha_n$ is slightly smaller than (resp. slightly larger than) $\alpha_n^*$ then the probability that $\mathbb{G}(n; \pmb{\mu},\pmb{K},P,\alpha)$ is connected is close to zero (resp. close to one); e.g., see \cite[Section 7.5]{Bollobas}.
Our results  constitute an asymptotic solution of the network reliability problem for inhomogeneous random key graphs. More specifically, we show that $\alpha_n$ exhibits a threshold behavior as given at (\ref{scaling_condition_KG}). Although asymptotic in nature, these results can still provide useful insights about the reliability of heterogeneous WSNs with number of sensors $n$ being on the order of hundreds; see Section \ref{sec:numerical} for numerical experiments.

\section{Numerical Results}
\label{sec:numerical}
\begin{figure}[t]
\centerline{\includegraphics[scale=0.3]{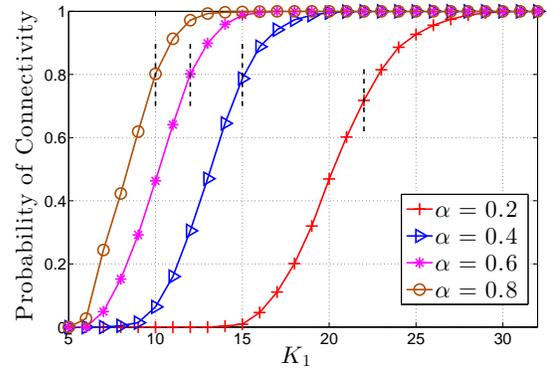}}
\caption{Empirical probability that $\mathbb{G}(n;\pmb{\mu},\pmb{K}, P, \alpha)$ is connected 
with $n = 500$, $\pmb{\mu}=(1/4,1/4,1/4,1/4)$,   $\pmb{K}=(K_1,K_1+5,K_1+10,K_1+15)$, and
$P = 10^4$.
Vertical dashed lines give 
the minimum $K_1$ for which (\ref{eq:numerical_critical}) holds.}
\label{fig:1}
\end{figure}

We  present numerical results that support Theorem~\ref{theorem:isolated_nodes} and Theorem~\ref{theorem:connectivitiy} in the finite node regime. In all experiments, we fix the number of nodes at $n = 500$ and the size of the key pool at $P = 10^4$. To help better visualize the results,  we use the curve fitting tool of MATLAB. In Figure~\ref{fig:1}, we consider the link-failure parameters $\alpha = 0.2$, $\alpha = 0.4$, $\alpha = 0.6$, and $\alpha = 0.8$, while varying the parameter $K_1$ (i.e., the smallest key ring size) from $5$ to $35$. The number of classes is fixed at $4$ with $\pmb{\mu}=\{0.25,0.25,0.25,0.25\}$ and we set $K_2=K_1+5$, $K_3=K_1+10$, and $K_4=K_1+15$. For each parameter pair $(\pmb{K}, \alpha)$, we generate $200$ independent samples of the graph $\mathbb{ G}(n;\pmb{\mu},\pmb{\Theta})$ and count the number of times (out of a possible 200) that the obtained graphs i) have no isolated nodes and ii) are connected. Dividing the counts by $200$, we obtain the (empirical) probabilities for the events of interest.  We observed that $\mathbb{ G}(n;\pmb{\mu},\pmb{\Theta})$ is connected whenever it has no isolated nodes yielding the same empirical probability for both events. This is in parallel with the asymptotic equivalence of the two properties as implied by Theorems \ref{theorem:isolated_nodes} and \ref{theorem:connectivitiy}.

In Figure~\ref{fig:1} we show the {\em critical} threshold of connectivity \lq\lq predicted" by Theorem~\ref{theorem:connectivitiy} by a vertical dashed line. More specifically, the vertical dashed lines stand for the minimum integer $K_1$ such that 
\begin{equation}
\lambda_1(n)=\sum_{j=1}^4 \mu_j \left( 1- \frac{\binom{P-K_j}{K_1}}{\binom{P}{K_1}} \right) > \frac{1}{\alpha} \frac{\log n}{n}.
\label{eq:numerical_critical}
\end{equation}
We see that the probability of connectivity transitions from zero to one within relatively small variation of $K_1$, with critical values of $K_1$ from (\ref{eq:numerical_critical}) lying within this transition interval.

\begin{figure}[t]
\centerline{\includegraphics[scale=0.3]{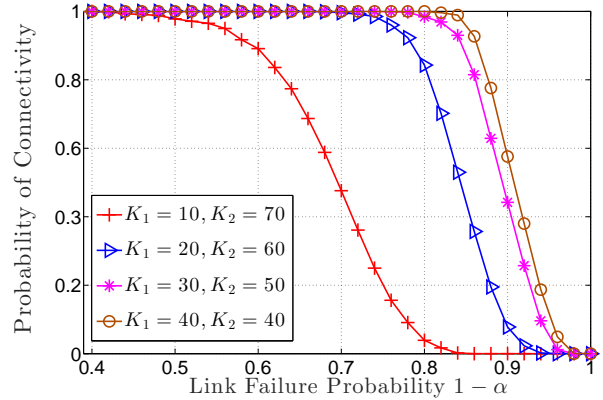}}
\caption{ 
 Empirical probability that $\mathbb{ G}(n;\pmb{\mu},\pmb{K},P,\alpha)$ is connected with $n = 500$, $\pmb{\mu}=(1/2,1/2)$, and $P = 10^4$;
 we consider four choices of $\pmb{K}=(K_1,K_2)$ each with the same mean.
Vertical dashed lines give the minimum $\alpha$ for which (\ref{eq:numerical_critical}) holds.}
\label{fig:5}
\end{figure}

Figure~\ref{fig:5} is generated in a similar manner with Figure~\ref{fig:1}, this time with an eye towards understanding the impact of the minimum key ring size $K_1$ on network reliability. To that, we fix the number of classes at $2$ with $\pmb{\mu}=\{0.5,0.5\}$ and consider 
four different key ring sizes $\pmb{K}$
each with mean 30; namely, we consider
$\pmb{K} = \{10,70\}$, $\pmb{K} = \{20,60\}$, $\pmb{K} = \{30,50\}$, and $\pmb{K} = \{40,40\}$.
As we compare the probability of connectivity in the resulting networks with link failure probability ranging from zero to one, we see that network reliability improves dramatically as the minimum key ring size $K_1$ increases.

\section{Proof of Theorem~\ref{theorem:isolated_nodes}}
\label{sec:proof_isolated}
The proof of Theorem~\ref{theorem:isolated_nodes} relies on the method of first and second moments applied to the number of isolated nodes in $\mathbb{G}(n;\pmb{\mu},\pmb{\Theta}_n)$. Let $I_n(\pmb{\mu},\pmb{\Theta}_n)$ denote the total number of isolated nodes in $\mathbb{G}(n;\pmb{\mu},\pmb{\Theta}_n)$, namely,
\begin{equation}
I_n(\pmb{\mu},\pmb{\Theta}_n)=\sum_{\ell=1}^n \pmb{1}[v_\ell \text{ is isolated in }\mathbb{G}(n;\pmb{\mu},\pmb{\Theta}_n)]
\label{eq:isolated_In}
\end{equation}

The method of first moment \cite[Eqn. (3.10), p.
55]{JansonLuczakRucinski} gives
\begin{equation} \nonumber
1-\mathbb{E}[I_n(\pmb{\mu},\pmb{\Theta}_n)]\leq \mathbb{P}[I_n(\pmb{\mu},\pmb{\Theta}_n)=0] 
\end{equation}

\subsection{Preliminaries}

Several technical results are collected here for convenience.
The first result is established in {\cite[Proposition~4.1]{Yagan/Inhomogeneous}} and follows easily from the scaling condition (\ref{scaling_condition_K}): For any scaling $K_1,K_2,\ldots,K_r,P:\mathbb{N}_0 \rightarrow \mathbb{N}_0^{r+1}$, we have
\begin{equation}
\lambda_1(n) \leq \lambda_2(n) \leq \ldots \leq \lambda_r(n)
\label{eq:isolated_ordering_of_lambda}
\end{equation}
for each $n=2,3,\ldots$. 

%
%

Another useful bound that will be used throughout is
\begin{align}
& (1 \pm x) \leq e^{\pm x}, \quad x \in (0,1)
\label{eq:isolated:exp_bound}
\end{align}

Finally, we find it useful to write
\begin{equation}
\log (1-x)=-x-\Psi(x)
\label{eq:isolated_log_decomp}
\end{equation}
where  $\Psi(x)=\int_{0}^{x} \frac{t}{1-t} \ \text{dt}$.
From L'H\^{o}pital's Rule, we have
\begin{equation}
\lim_{x\to 0}  \frac{\Psi(x)}{x^2}=\frac{-x-\log (1-x)}{x^2}=\frac{1}{2}.
\label{eq:isolated_hopital}
\end{equation}

\subsection{Establishing the one-law}
It is clear that in order to establish the one-law, namely that $ \lim_{n \to \infty} \mathbb{P}\left[  I_n(\pmb{\mu},\pmb{\Theta_n})=0\right]$, we need to show that
$\lim_{n \to \infty} \mathbb{E}[I_n(\pmb{\mu},\pmb{\Theta}_n)]=0$. 
Recalling (\ref{eq:isolated_In}), we have
\begin{align}
&\mathbb{E}\left[I_n(\pmb{\mu},\pmb{\Theta}_n)\right] \nonumber \\
&=n \sum_{i=1}^r \mu_i \mathbb{P}\left[v_1 \text{ is isolated in }\mathbb{G}(n;\pmb{\mu},\pmb{\Theta}_n) \given[\big] t_1=i\right]\nonumber \\
&=n \sum_{i=1}^r \mu_i \mathbb{P}\left[\cap_{j=2}^n [v_j \nsim v_1] \given[\big] t_1=i\right]\nonumber \\
&=n \sum_{i=1}^r \mu_i \left(\mathbb{P}\left[v_2 \nsim v_1 \given[\big] t_1=i\right]\right)^{n-1} \label{eq:isolated_independence}
\end{align}
where (\ref{eq:isolated_independence}) follows by the independence of the rvs $\{v_j \nsim v_1\}_{j=1}^n$ given $\Sigma_1$. By conditioning on the class of $v_2$, we find
\begin{align}
\mathbb{P}[v_2 \nsim v_1 \given[\big] t_1=i]
&=\sum_{j=1}^r \mu_j \mathbb{P}[v_2 \nsim v_1 \given[\big] t_1=i,t_2=j]\nonumber \\
&=\sum_{j=1}^r \mu_j (1-\alpha p_{ij})=1-\Lambda_i(n).
\label{eq:isolated_OneLaw_first_step}
\end{align}
Using (\ref{eq:isolated_OneLaw_first_step}) in (\ref{eq:isolated_independence}), and recalling (\ref{eq:isolated_ordering_of_lambda}),  (\ref{eq:isolated:exp_bound}) we obtain
\begin{align*}
\mathbb{E}[I_n(\pmb{\mu},\pmb{\Theta}_n)] &= n \sum_{i=1}^r \mu_i \left(1-\Lambda_i(n)\right)^{n-1}\nonumber \\
&\leq n \left(1-\Lambda_1(n)\right)^{n-1}\nonumber \\
&= n \left(1-c_n \frac{\log n}{n}\right)^{n-1}\nonumber \\
&\leq  e^{\log n \left(1-c_n  \frac{n-1}{n}\right)}
\end{align*}
Taking the limit as $n$ goes to infinity, we immediately get
$\lim_{n \to \infty} \mathbb{E}[I_n(\pmb{\mu},\pmb{\Theta}_n)]=0$
since $\lim_{n \to \infty} (1-c_n  \frac{n-1}{n})=1-c < 0$ under the enforced assumptions (with $c>1$) and the one-law is established.

\subsection{Establishing the zero-law}
Our approach in establishing the zero-law relies on the method of second moment applied to a variable that counts the number of nodes that are class-$1$ and isolated. Clearly if we can show that whp there exists at least one class-$1$ node that is isolated under the enforced assumptions (with $c<1$) then the zero-law would immediately follow.

Let $Y_n(\pmb{\mu},\pmb{\Theta}_n)$ denote the number of nodes that are class-$1$ and isolated in $\mathbb{G}(n;\pmb{\mu},\pmb{\Theta}_n)$, and let
\begin{equation} \nonumber
x_{n,i}(\pmb{\mu},\pmb{\Theta}_n)=\pmb{1}[t_i=1 \cap v_i \text{ is isolated in }\mathbb{G}(n;\pmb{\mu},\pmb{\Theta}_n)],
\end{equation}
then we have $Y_n(\pmb{\mu},\pmb{\Theta}_n)=\sum_{i=1}^n x_{n,i}(\pmb{\mu},\pmb{\Theta}_n)$. By applying the method of second moments
\cite[Remark 3.1, p. 55]{JansonLuczakRucinski}  on $Y_n(\pmb{\mu},\pmb{\Theta}_n)$, we get
\begin{equation}
\mathbb{P}[Y_n(\pmb{\mu},\pmb{\Theta}_n)=0] \leq 1-\frac{\mathbb{E}[Y_n(\pmb{\mu},\pmb{\Theta}_n)]^2}{\mathbb{E}[Y_n(\pmb{\mu},\pmb{\Theta}_n)^2]}  
\label{eq:isolated_ZeroLaw_bound}
\end{equation}
where
\begin{equation}
\mathbb{E}[Y_n(\pmb{\mu},\pmb{\Theta}_n)]=n \mathbb{E}[x_{n,1}(\pmb{\mu},\pmb{\Theta}_n)]
\label{eq:isolated_ZeroLaw_first_part}
\end{equation}
and
\begin{align}
\begin{split}
\mathbb{E}[Y_n(\pmb{\mu},\pmb{\Theta}_n)^2]=&n \mathbb{E}[x_{n,1}(\pmb{\mu},\pmb{\Theta}_n)]\\
&+n(n-1)\mathbb{E}[x_{n,1}(\pmb{\mu},\pmb{\Theta}_n) x_{n,2}(\pmb{\mu},\pmb{\Theta}_n)]
\end{split}
\label{eq:isolated_ZeroLaw_second_part}
\end{align}
by exchangeability and the binary nature of the rvs $\{x_{n,i}(\pmb{\mu},\pmb{\Theta}_n) \}_{i=1}^n$.
Using (\ref{eq:isolated_ZeroLaw_first_part}) and  (\ref{eq:isolated_ZeroLaw_second_part}), we get
\begin{equation} \nonumber
\begin{split}
\frac{\mathbb{E}[Y_n(\pmb{\mu},\pmb{\Theta}_n)^2]}{\mathbb{E}[Y_n(\pmb{\mu},\pmb{\Theta}_n)]^2}  =& \frac{1}{n \mathbb{E}[x_{n,1}(\pmb{\mu},\pmb{\Theta}_n)]} \\
&+ {\frac{n-1}{n} \frac{\mathbb{E}[x_{n,1}(\pmb{\mu},\pmb{\Theta}_n) x_{n,2}(\pmb{\mu},\pmb{\Theta}_n)]}{\mathbb{E}[x_{n,1}(\pmb{\mu},\pmb{\Theta}_n)]^2}}
\end{split}
\end{equation}

Accordingly, in order to establish the zero-law, we need to show that
\begin{equation}
\lim_{n \to \infty} n \mathbb{E}[x_{n,1}(\pmb{\mu},\pmb{\Theta}_n)]= \infty,
\label{eq:isolated_ZeroLaw_first_condition}
\end{equation}
and
\begin{equation}
\limsup_{n \to \infty} \left(\frac{\mathbb{E}[x_{n,1}(\pmb{\mu},\pmb{\Theta}_n) x_{n,2}(\pmb{\mu},\pmb{\Theta}_n)]}{\mathbb{E}[x_{n,1}(\pmb{\mu},\pmb{\Theta}_n)]^2}\right) \leq 1.
\label{eq:isolated_ZeroLaw_second_condition}
\end{equation}

The following propositions establish (\ref{eq:isolated_ZeroLaw_first_condition}) and (\ref{eq:isolated_ZeroLaw_second_condition}) which in turn establish the zero-law.

{\prop
Consider a scaling $K_1,\ldots,K_r,P:\mathbb{N}_0 \rightarrow \mathbb{N}_0^{r+1}$ and a scaling $\alpha:\mathbb{N}_0 \rightarrow (0,1)$ such that (\ref{scaling_condition_KG}) holds with $\lim_{n \to \infty} c_n=c>0$. Then, we have
\begin{equation*}
\lim_{n \to \infty} n \mathbb{E}[x_{n,1}(\pmb{\mu},\pmb{\Theta}_n)]= \infty, \quad \text{if } c<1
\end{equation*}
}
\myproof
We have
\begin{align}
&n \mathbb{E}\left[x_{n,1}(\pmb{\mu},\pmb{\Theta}_n)\right] \nonumber \\
&=n \mu_1 \mathbb{P}\left[\cap_{j=2}^n [v_j \nsim v_1] \given[\big] t_1=1\right]\nonumber \\
&=n \mu_1 \left(\sum_{j=1}^r \mu_j \mathbb{P}\left[v_2 \nsim v_1 \given[\big] t_1=1,t_2=j\right]\right)^{n-1}\nonumber \\
&=n \mu_1 \left(\sum_{j=1}^r \mu_j (1-\alpha_n p_{1j})\right)^{n-1} \label{eq:int_isol_prob_osy} \\
&=n \mu_1 \left(1-\Lambda_1(n)\right)^{n-1} = \mu_1 e^{\beta_n} 
\label{eq:isolated_ZeroLaw_simp1}
\end{align}
where
\begin{align}
\beta_n&=\log n+(n-1)\log (1-\Lambda_1(n)) \nonumber \\
&=\log n-(n-1)\left(\Lambda_1(n)+\Psi(\Lambda_1(n))\right)\nonumber \\
&=\log n-(n-1)\left(c_n \frac{\log n}{n}+\Psi \left(c_n \frac{\log n}{n}\right)\right)\nonumber \\
&=\log n \left(1-c_n \frac{n-1}{n}\right) \nonumber \\
& \quad -(n-1) \left(c_n \frac{\log n}{n} \right)^2 \frac{\Psi \left(c_n \frac{\log n}{n}\right)}{\left(c_n \frac{\log n}{n}\right)^2}
 \label{eq:isolated_ZeroLaw_simp2}
\end{align}
by virtue of (\ref{eq:isolated_log_decomp}). Now, recalling (\ref{eq:isolated_hopital}), we have
\begin{equation}
\lim_{n \to \infty} \frac{\Psi \left(c_n \frac{\log n}{n}\right)}{\left(c_n \frac{\log n}{n}\right)^2} = \frac{1}{2}
\label{eq:isolated_ZeroLaw_simp3}
\end{equation}
since $c_n \frac{\log n}{n}=o(1)$. Thus, $\beta_n=\log n \left(1-c_n \frac{n-1}{n}\right)-o(1)$.
Using (\ref{eq:isolated_ZeroLaw_simp1}), (\ref{eq:isolated_ZeroLaw_simp2}), (\ref{eq:isolated_ZeroLaw_simp3}), and letting $n$ go to infinity, we get
\begin{equation*}
\lim_{n \to \infty} n \mathbb{E}[x_{n,1}(\pmb{\mu},\pmb{\Theta}_n)]= \infty
\end{equation*}
whenever $\lim_{n \to \infty} c_n=c < 1$.
\myendpf

{\prop
Consider a scaling $K_1,\ldots,K_r,P:\mathbb{N}_0 \rightarrow \mathbb{N}_0^{r+1}$ and a scaling $\alpha:\mathbb{N}_0 \rightarrow (0,1)$ such that (\ref{scaling_condition_KG}) holds with $\lim_{n \to \infty} c_n=c>0$. Then, we have (\ref{eq:isolated_ZeroLaw_second_condition}) if $c<1$.
\label{prop:new_osy}
}

We omit the proof of Proposition~\ref{prop:new_osy} from this conference version. All details can be found in \cite{Rashad/Inhomo}.

\section{Proof of Theorem~\ref{theorem:connectivitiy}}
The proof of Theorem~\ref{theorem:connectivitiy} is lengthy and technically involved. Therefore, we omit most of the details in this conference version. All details can be found in \cite{Rashad/Inhomo}. In this section, we present an outline of our proof. 
Let $C_n(\pmb{\mu},\pmb{\Theta}_n)$ denote the event that the graph $\mathbb{G}(n,\pmb{\mu},\pmb{\Theta}_n)$ is connected,
and with a slight abuse of notation, let $I_n(\pmb{\mu},\pmb{\Theta}_n)$ denote the event that the graph $\mathbb{G}(n,\pmb{\mu},\pmb{\Theta}_n)$ has no isolated nodes.
Clearly, if a random graph is connected then it does not have any isolated node, hence
$
C_n(\pmb{\mu},\pmb{\Theta}_n) \subseteq I_n(\pmb{\mu},\pmb{\Theta}_n)
$
and we get
\begin{equation}
\mathbb{P}[C_n(\pmb{\mu},\pmb{\Theta}_n)] \leq \mathbb{P}[I_n(\pmb{\mu},\pmb{\Theta}_n)]
\label{eq:conn_ZeroLaw}
\end{equation}
and
\begin{align} \label{eq:conn_OneLaw}
\mathbb{P}[C_n(\pmb{\mu},\pmb{\Theta}_n)^c] & = \mathbb{P}[I_n(\pmb{\mu},\pmb{\Theta}_n)^c]+\mathbb{P}[C_n(\pmb{\mu},\pmb{\Theta}_n)^c \cap I_n(\pmb{\mu},\pmb{\Theta}_n)].
\end{align}

In view of (\ref{eq:conn_ZeroLaw}), we obtain the zero-law for connectivity, i.e., that
\begin{equation} \nonumber
\lim_{n\to\infty} \mathbb{P}[\mathbb{G}(n;\pmb{\mu},\pmb{\Theta}_n) \text{ is connected}]= 0    \quad \text{ if } \quad c<1,
\end{equation}
immediately from the zero-law part of
Theorem \ref{theorem:isolated_nodes}, i.e., from that 
$\lim_{n\to\infty} \mathbb{P}[I_n(\pmb{\mu},\pmb{\Theta}_n)]=0$ if $c<1$.
It remains to establish the one-law for connectivity. 
From Theorem \ref{theorem:isolated_nodes} and (\ref{eq:conn_OneLaw}), we see that the one-law for connectivity, i.e., that
\[
\lim_{n\to\infty} \mathbb{P}[\mathbb{G}(n;\pmb{\mu},\pmb{\Theta}_n) \text{ is connected}]= 1    \quad \text{ if } \quad c > 1,
\]
will follow if we show that
\begin{align}
\lim_{n \to \infty} \mathbb{P}[C_n(\pmb{\mu},\pmb{\Theta}_n)^c \cap I_n(\pmb{\mu},\pmb{\Theta}_n)] = 0.
\label{eq:conn_bounding}
\end{align}

The proof of the one-law passes through obtaining a proper upper bound for  (\ref{eq:conn_bounding}) and then showing that the bound goes to zero as $n$ gets to infinity (with $c>1$) under appropriate conditions
of the parameter scalings. Due to space limitations, the details of this technically involved result are given in \cite{Rashad/Inhomo}.

\section*{Acknowledgment}
This work has been supported in part by National Science Foundation through grants CCF-1617934 and CCF-1422165
and in part by the start-up funds from the Department of Electrical and Computer Engineering at Carnegie Mellon University. 
\ifCLASSOPTIONcaptionsoff
  \newpage
\fi

\bibliographystyle{IEEEtran}

\bibliography{IEEEabrv,CDC}

\end{document}